# Evidence of Open Access of scientific publications in Google Scholar: a large-scale analysis

Alberto Martín-Martín[1], Rodrigo Costas[2,3], Thed van Leeuwen[2], Emilio Delgado López-Cózar[1]

## Abstract

This article uses Google Scholar (GS) as a source of data to analyse Open Access (OA) levels across all countries and fields of research. All articles and reviews with a DOI and published in 2009 or 2014 and covered by the three main citation indexes in the Web of Science (2,269,022 documents) were selected for study. The links to freely available versions of these documents displayed in GS were collected. To differentiate between more reliable (sustainable and legal) forms of access and less reliable ones, the data extracted from GS was combined with information available in DOAJ, CrossRef, OpenDOAR, and ROAR. This allowed us to distinguish the percentage of documents in our sample that are made OA by the publisher (23.1%, including Gold, Hybrid, Delayed, and Bronze OA) from those available as Green OA (17.6%), and those available from other sources (40.6%, mainly due to ResearchGate). The data shows an overall free availability of 54.6%, with important differences at the country and subject category levels. The data extracted from GS yielded very similar results to those found by other studies that analysed similar samples of documents, but employed different methods to find evidence of OA, thus suggesting a relative consistency among methods.

---

[1] Facultad de Comunicación y Documentación, Universidad de Granada, Granada, Spain.
[2] CWTS, Leiden University, Leiden, The Netherlands.
[3] DST-NRF Centre of Excellence in Scientometrics and Science, Technology and Innovation Policy, Stellenbosch University, South Africa.

✉ Alberto Martín-Martín
   albertomartin@ugr.es

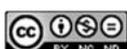

# 1. Introduction

## 1.1. Beginnings of the Open Access movement

The widespread adoption of web technologies removed most of the physical impediments for accessing scientific information (Harnad, 2001). Since then, the issue of Open Access (henceforth referred to as OA) to the scholarly literature has been hotly debated by all sorts of actors in the academic community, including researchers, publishers, funding institutions, librarians, and policy makers. Many of these discussions revolved around the ways in which the system of scholarly communication should change, taking advantage of this new virtual environment to become more effective and efficient and thus hopefully solve problems like the affordability and accessibility to scientific information that afflict many research institutions.

One of the first crystallizations of these intentions to change the scholarly communication system was the Budapest Open Access Initiative (Chan et al., 2002) (BOAI). This was the first time the term "Open Access" was used, although the practices described in that document had already been taking place in some scientific communities long before that date. The BOAI defined OA to the literature as:

"*free availability on the public internet, permitting any users to read, download, copy, distribute, print, search, or link to the full texts of these articles, crawl them for indexing, pass them as data to software, or use them for any other lawful purpose, without financial, legal, or technical barriers other than those inseparable from gaining access to the internet itself. The only constraint on reproduction and distribution, and the only role for copyright in this domain, should be to give authors control over the integrity of their work and the right to be properly acknowledged and cited*".

Additionally, the BOAI also described the two main ways to realise the goal of OA: by self-archiving documents in public archives (which later came to be known as Green OA), or by publishing in OA journals (later dubbed as Gold OA). Poynder (2018) provides a historic overview of OA since the BOAI declaration.

Since the original BOAI declaration was first published, the discussion has continued and the panorama of scholarly publishing and OA has greatly changed. All actors have had to adapt in some way to the new reality. In addition, the Web gave rise to new types of academic platforms, which further complicated the issue of access to scientific information by expanding the access points to scientific content (e.g. Google Scholar, ResearchGate, etc.). Some of these new platforms were quickly adopted by the scientific community and have already become an important part of the system. These platforms will be discussed later on.

## 1.2. Reactions of academic institutions, funders and publishers to OA

In the beginnings of the OA movement, a great emphasis was put on the importance of authors self-archiving their own publications on public repositories (Harnad, 2001). Many research institutions, which saw in self-archiving a potential solution to the *journal affordability problem* (the problem of selecting which journals to subscribe to, when economic resources are limited), put systems in place to allow researchers to self-archive and make public their research. These institutional repositories are under the direct control of the institution, and are usually managed by the libraries. Additionally, other subject-specific repositories were launched. Apart from arXiv[4], the physics repository created in 1991 in Cornell University, many other repositories are now available to researchers. ROAR[5] (Registry of Open Access Repositories) and OpenDOAR[6] (Directory of Open Access Repositories) provide an exhaustive list of these institutional and subject based OA repositories. More recently, there has been an explosion in the growth of the so-called preprint servers, largely enabled by the infrastructure developed by the Open Science Framework[7], a project launched by the Center for Open Science, which is a non-profit organization founded in 2013 to "increase the openness, integrity, and reproducibility of scientific research" (Mellor, 2016, para. 6). These servers are designed to share manuscripts that still have not gone through a process of peer review, although they usually welcome accepted manuscripts as well.

One of the notions that has served to justify the need of OA is that money from public institutions to fund research was not realizing its true potential, because most publicly-funded research ended up behind publishers' paywalls, and other researchers who could make use of that research had no access to it. For these reasons, many funding institutions, governments, and policy makers started to issue OA mandates to force researchers who use their funding to make their results OA. Among these we can find the National Institutes of Health (NIH) in the USA, the Research Councils in the United Kingdom, or the European Research Council. In 2016, the European Union announced its resolve to make all scientific publications based on publicly-funded research freely accessible by 2020 (Enserink, 2016). ROARMAP[8] (Registry of Open Access Repository Mandates and Policies) provides an exhaustive database of OA mandates issued by all kinds of organizations worldwide.

Largely because of these mandates, most publishers adapted their business models, which previously relied almost exclusively on journal subscriptions paid by academic institutions, to business models compatible with the OA requirements mandated by funders:
- Gold OA journals publish all their articles as OA. Their revenue usually comes from charging Article Processing Charges (APC) to authors instead of charging subscription fees to academic libraries. There is much controversy concerning the price of these APCs, which range from a few hundreds of dollars, to over $5,000 per article. There are also Gold OA journals that do not charge APCs to authors, and instead absorb publishing costs in other ways (like via member subscriptions fees in the cases of academic societies that also publish journals). These are sometimes called Diamond OA or Platinum OA journals (Fuchs & Sandoval, 2013; Haschak, 2007).
- Hybrid OA journals maintain the subscription model, but give authors the choice to make their article OA, also by paying an APC (Prosser, 2003; Walker, 1998). This model has also been controversial, because in addition to charging APCs to authors to make the articles OA, they still charge libraries ever-increasing subscription costs for access to the entire collection of articles published by the journals. This phenomenon has been dubbed "double-dipping", because publishers seem to be charging twice for the same content. Some publishers, like Elsevier, claim that Hybrid OA articles are excluded when calculating subscription costs[9], while other publishers compensate institutions "for the extra money they are putting into the system through payment of APCs" (Kingsley, 2017, para. 3) by means of the so-called "offset agreements", which can take many forms. Lawson (2018) reports on the offset agreements made with publishers by the organization JISC Collections, which works on behalf of UK academic libraries.
- <u>Delayed OA journals</u> are subscription journals that convert their articles to OA once a specific amount of time has passed after publication. Laakso and Björk (2013) analyzed a sample of 111,312 articles published in 492 journals and found that 77.8% of them were available from the publisher website twelve months after publication. The percentage reached 85.4% 24 months after publication.
- <u>Gratis Access Journals</u> (Suber, 2008a, 2008b): journals that make their articles free-to-read, but don't extend other rights to users (such as reuse or distribution) apart from the right to read. The publisher retains the copyright of these articles. This type of access is also referred to as "public access", especially by the publishing industry (Crotty, 2017). Sometimes publishers intend to maintain access to these documents free indefinitely, but sometimes access is only free for a specific period of time (promotional access). Therefore, this type should not be conflated with Gold, Hybrid, or Delayed OA.

---

[4] https://arxiv.org/
[5] http://roar.eprints.org/
[6] http://www.opendoar.org/
[7] https://osf.io/
[8] http://roarmap.eprints.org/

The costs of subscriptions and APCs are continually increasing (Tickell et al., 2017). This fact has led a number of institutions and governments to re-negotiate the so-called Big Deals (flat rates to access large numbers of journals published by a single publisher) so that they also include flat rates or considerable discounts for the APCs of the articles their researchers publish (Elsevier, 2015). In other cases, governments have refused to pay the increasing costs that large commercial publishers demanded. This was the case with Germany and the publisher Elsevier. A coalition of German institutions (grouped under the name project DEAL[10]) decided not to renew their license to Elsevier content at the end of 2016. Elsevier subsequently stopped allowing them to access its content, but decided to restore access shortly after, "in good faith" while negotiations lasted. By June 2018 an agreement had still not been reached. After Germany, other countries have followed suit: in March 2018, the Couperin consortia in France decided to not to renew their agreement with Springer-Nature, and in May 2018 the Bibsam Consortium in Sweden decided not to renew their agreement with Elsevier (Else, 2018).

Most journal publishers also offer alternative sharing policies for the articles that they do not publish as OA. The freedom these policies give to researchers to self-archive their content greatly varies by publisher and by specific journal. These policies often include embargo periods that prohibit authors to share their research on public repositories for a period of time after publication (from less than a year, to over two years). Despite initiatives like

---

[9] https://www.elsevier.com/about/our-business/policies/pricing#dipping
[10] https://www.projekt-deal.de/about-deal/

Sherpa/Romeo[11] or the publisher-backed How Can I Share It[12], which try to aggregate and standardise publisher's sharing policies, it is difficult to keep track of them because they change over time, usually to become more restrictive regarding how, where, and when self-archiving is permitted (Gadd & Troll Covey, 2016; Kingsley, 2013). These policies are often arbitrary and complicated, for example allowing to share an article immediately upon publication from the author's personal website, but imposing an embargo to share the same article from an institutional repository (Bolick, 2017; Tickell et al., 2017).

## 1.3. New players in the system

Other types of platforms, different from repositories and publishers but also with a large impact in the free availability of scholarly literature, have been launched since the BOAI declaration. In 2007 the academic search engine CiteSeerX[13] (based on an even earlier version called CiteSeer) was launched by Pennsylvania State University. In 2008, the academic social networks ResearchGate[14] and Academia.edu[15] were launched. In 2015, the search engine Semantic Scholar[16], developed by the Allen Institute for Artificial Intelligence, was launched, focusing mostly on the areas of Computer Science, and recently also Biomedicine. All these platforms share the characteristic that they host copies of the full texts of scholarly documents (automatically harvested from other sources or uploaded by users themselves) and make them available to their users, thus becoming another source from which readers can access scientific information.

Academic social networks (ASN) in particular have attracted a lot of attention because of how quickly users have taken to sharing their work on them (Björk, 2016). Borrego (2017) found that researchers from 13 Spanish universities used ResearchGate much more frequently to upload and share their research than the repositories available at their institutions. Martín-Martín, Orduna-Malea, Ayllón, and Delgado López-Cózar (2014; 2016), and Jamali and Navabi (2015) studied the free accessibility to a sample of documents covered by Google Scholar. Both studies found that ResearchGate was the source that provided the highest number of freely available full texts. However, full text documents in ASNs are uploaded by researchers themselves and, unlike OA repositories, these platforms do not carry out any kind of checks to guarantee copyright compliance. This resulted in a large portion of documents being accessible from ASN in violation of their copyright. Jamali (2017) found that 51.3% of the non-OA documents in a sample of 500 random documents were available from ResearchGate in violation of their copyright.

Moreover, despite some similarities, academic social networks engage in practices that clearly set them apart from OA repositories. The ongoing dispute between publishers and ResearchGate (Coalition for Responsible Sharing, 2017a, 2017b, 2017c) is unequivocal proof of the instability of these platforms as sources of full texts. A related issue is that in ResearchGate users are allowed to delete full texts of documents they have uploaded, even in the cases when the platform generates a DOI for the document (through their collaboration with DataCite[17]). This entirely differs from the policies of repositories such as arXiv or socArxiv, where the academic record is always maintained (authors cannot delete files but retain the right to issue a retraction notice if they feel a document they deposited should no longer be used). Full texts uploaded by a user to ResearchGate are also deleted if the user deletes his/her account in the platform. Academia.edu also engages in practices that make it different from repositories. This academic social network requires users to log in to their platform to access full texts. However, perhaps because this contravenes Google Scholar's indexing policies[18], they left open a back door so that users coming from a Google Scholar search[19] would be allowed to access full texts without the need to log in. Presumably, they did this to avoid being dropped as a source by Google Scholar, a large source of web traffic given its huge user-base. These cases raise the need to distinguish between merely uploading a document to the Internet (to ResearchGate, Academia.edu or to any privately managed personal website) and depositing or archiving a document in a repository, which usually provides more guarantees as to the long-term preservation of the documents that they host.

---

[11] http://www.sherpa.ac.uk/romeo
[12] http://www.howcanishareit.com/
[13] http://citeseerx.ist.psu.edu
[14] https://www.researchgate.net
[15] http://www.academia.edu
[16] https://www.semanticscholar.org

There is another player who is currently having a major influence in the accessibility to scholarly literature: Sci-Hub. This website was launched in 2011 by a graduate student called Alexandra Elbakyan, and it illegally provides access to over 60 million research articles. Elbakyan developed a system that automatically accesses publisher websites using credentials *donated* by users who work at institutions with access to paywalled journal articles. There are reports, however, that claim that some of these credentials might have been stolen rather than donated (Bohannon, 2016). The system then copies the full texts of articles to the Library Genesis database (LibGen), which is the platform that hosts the articles that in turn are provided to the users. The kind of copyright-infringing access that Sci-Hub provides is sometimes called Robin Hood OA, Rogue OA, and Black OA (Archambault et al., 2014; Björk, 2017; Green, 2017). Despite the efforts made by large commercial publishers like Elsevier to shut down Sci-Hub's operations, the website remained functional at the time of this writing, providing access to the vast majority of recently-published paywalled articles (Himmelstein et al., 2018) and virtually providing access to all scientific publications worldwide.

## 1.4. Current landscape of free availability of scientific information

To summarise the scenario described above, Figure 1 provides a representation of the main paths by which a journal article may become freely available on the Web.

---

[17] https://www.datacite.org/
[18] https://scholar.google.com/intl/en/scholar/inclusion.html#content
[19] Technically speaking, users who accessed Academia.edu with the Referer HTML request header "https://scholar.google.com"

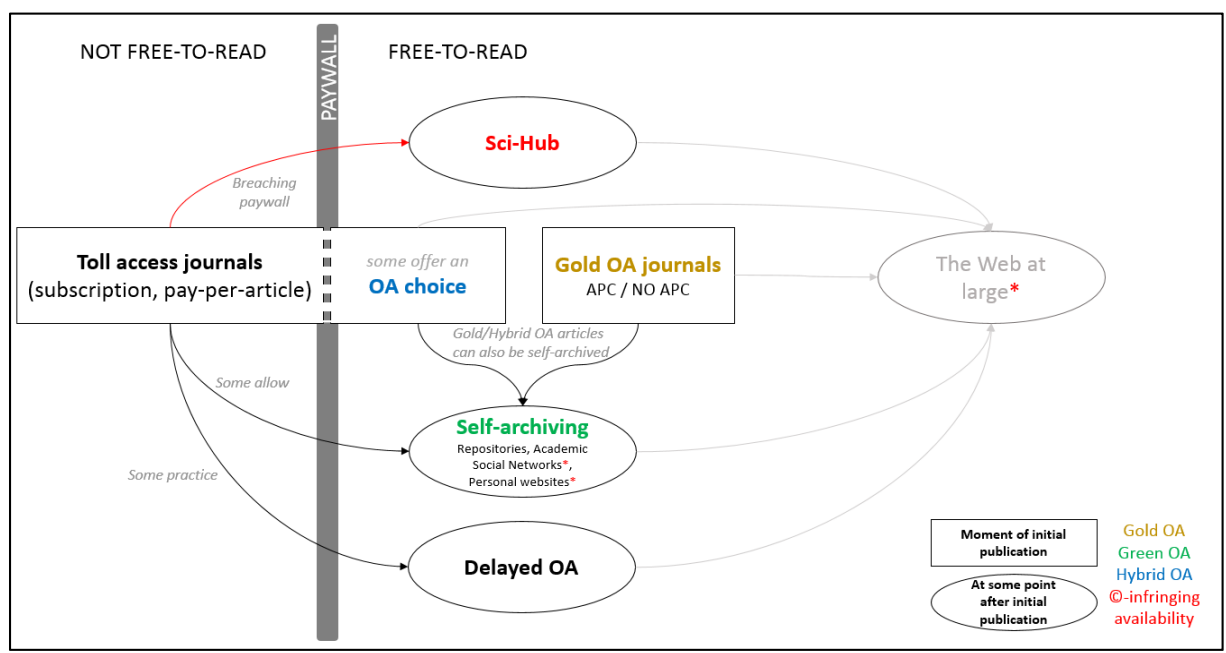

*Figure 1. Model of free availability of academic journal articles: Where are freely available journal articles hosted?*

The figure divides articles in two different spaces: the space in which articles are *not free-to-read* (to the left of the paywall) and the space in which articles are *free-to-read* (to the right of the paywall).

Articles published in Gold OA journals (regardless of whether they charge APCs or not), and articles published in OA in Hybrid journals are immediately made OA, hence their placement in the *free-to-read* section of Figure 1.

Articles that are initially *not free-to-read* (published in toll access journals) may become *free-to-read* in several ways (represented by lines going from the *toll access journals* box to the *free-to-read* space in Figure 1):
- By breaching the paywall, generating copyright-infringing availability (represented in the figure with a red continuous line and red asterisks). This is the case of Sci-hub, which cannot really be considered as a sustainable form of OA (van Leeuwen, Meijer, Yegros-Yegros, & Costas, 2017).
- Via self-archiving, when the journal allows it (represented by a line from the *toll access journals* box to the *free-to-read* space). Self-archiving mostly takes place in repositories, academic social networks, and personal websites. Repositories (both institutional or subject-specific) usually check for copyright compliance when articles are submitted. In personal websites and academic social networks, however, no such checks are made. Therefore, these venues might also contain articles in violation of their copyright (Jamali, 2017). This is represented with red asterisks in Figure 1. There is also a line from Gold and Hybrid OA journals to the self-archiving section, because OA articles can always be self-archived.
- Delayed OA, which is practiced by some journals (also represented by a line from the *toll access journals* box to the *free-to-read* section).

Once articles are *free-to-read* in any of the ways described above, they may be distributed (legally or not) to any other part of the Web at large. For example, some platforms, like the academic search engines CiteSeerX and Semantic Scholar harvest the full texts of articles available in other sources, and provide a copy from their own servers.

Lastly, apart from being freely available, documents must also be discoverable in order to be used. There are several services that address the *discoverability problem*, like the academic search engines BASE[20] and Google Scholar, or the browser extension Unpaywall[21]. Google Scholar and Unpaywall are described in more detail in the following section. Coverage of freely available documents varies by platform. Google Scholar, the focus of this article, serves as a gateway for all types of sources described in Figure 1, with the exception of Sci-Hub.

## 1.5. Quantification of OA levels

In a scenario like the one described above, it is not surprising that the question of how much of the scholarly literature is openly accessible (or at least freely available) has attracted much attention, because many agents of the scholarly community are interested in its answer. Funders are interested in the degree to which their OA mandates are being obeyed. Libraries need to decide how to best use their acquisitions budget (whether to renew, renegotiate, or cancel license agreements with publishers). Publishers routinely monitor how the documents they publish are shared on the Web in order to protect their business. Countries, for their part, want to know how much of the scientific literature published by its researchers is openly accessible. Researchers may also be interested in the proportion of their publications that is openly accessible, especially if this is an issue that is taken into account in the performance evaluations to which they are subjected in their country.

Numerous studies have analyzed the levels of OA for different samples of documents, presenting results at various levels of aggregation (publication year, subject areas, countries of authors' affiliations, OA types...). Methods to ascertain levels of OA include using data collected by custom crawlers (1science database, Unpaywall data) and carrying out searches in diverse search engines (BASE, Google, Google Scholar...). Table 1 contains information on the sample of documents analyzed, source of OA evidence used, and OA levels found by studies that used a source of OA evidence other than Google Scholar.

---

[20] https://www.base-search.net/
[21] http://unpaywall.org/

*Table 1. Studies that analyse OA levels using sources of OA evidence other than Google Scholar*

| Study | Sample of documents | | | | | OA evidence | | | | OA levels |
|---|---|---|---|---|---|---|---|---|---|---|
| | Source | Field | Pub. Year | Doc types | Size | Source | Date of data collection | Levels of aggregation | Methodological Observations | |
| Björk et al., 2010 | Scopus (random) | All fields | 2008 | Articles | 1,837 | Searches on Google | 2009/10 | Subject areas, OA types | | 20.4% freely accessible (8.5% from publisher) |
| Gargouri, Larivière, Gingras, Carr, & Harnad, 2012 | Web of Science (random) | 11 fields | 1998-2006 | Articles | 110,212 | Custom crawler (no details given) | 2009 | Publication year, subject areas, OA types | | 20% freely accessible (average of entire period) |
| | = | 14 fields | 2005-2010 | = | 107,052 | = | 2011 | = | | 24% freely accessible (average of entire period). 21.4% as Green OA, 2.4% as Gold OA |
| Archambault et al., 2014 | Scopus (random) | All fields | 1996-2013 | Articles | ~ 245,000 | Custom crawler: Scielo, PubMed Central, ResearchGate, CiteSeerX, publisher websites, arXiv, repositories in ROAR and OpenDOAR | 2013/04, 2014/04 | OA types | Calibration factor (1.146) applied to account for limited recall of custom crawler | Over 50% of articles published 2007-2012 were freely available in 2014 |
| | Scopus (random) | 22 fields | 2008-2013 | Articles | ~ 1 million | = | 2014/04 | Subject areas, countries (ERA) | = | Top OA field (2011-2013): General Science & Technology (90%) Top OA countries (2008-2013): Netherlands, Croatia, Estonia, and Portugal (>70%) |
| van Leeuwen et al., 2017 | Web of Science (all records) | All fields | 2009-2014 | All types | Not declared | DOAJ, ROAD, CrossRef, PubMed Central, OpenAIRE | 2017 | Publication year, OA evidence source, countries | | Almost 30% of articles were OA. Top countries: Netherlands (37%), Sweden, Ireland, and UK (34%) |
| Smith et al., 2017 | PubMed (selected subject heading) | Global Health | 2010-2014 | Articles | 3,366 | PubMed, manual searches on Google | 2016 | OA types | | 29.2% OA from publisher, 27.2% Green OA, 1.3% OA from other sources. Total OA: ~ 58% |
| Science-Metrix Inc., 2018 | Web of Science (all records) | All fields | 2006-2015 | Not declared | Not declared | 1science database: scholarly material indexed in over 180,000 websites | 2016/07-09 | Publication year, countries, Subject areas, OA types | Calibration factor (1.2) applied. PubMed Central considered Gold OA; ResearchGate considered Green OA | Pub. Year 2006: 50%, pub. year 2011: 60% Top countries 2014: Brazil (74%), Netherlands (68%) Top fields: Health Sciences (59%) |
| Piwowar et al., 2018 | CrossRef (random) | All fields | All years | Articles | 100,000 | Unpaywall data | 2017/05 | Publication year, publisher, OA types | ResearchGate not included in Unpaywall | 27.9% are OA; 44.7% for pub. year 2015 |
| | Web of Science (random) | All fields | 2009-2015 | Articles and reviews | 100,000 | Unpaywall data | 2017/05 | Subject areas, OA types | = | 36.1% are OA |
| | Unpaywall use logs | All fields | All years | All types | 100,000 | Unpaywall data | 2017/06/05-11 | OA types | = | 47% of documents accessed by users via Unpaywall are OA |
| Bosman & Kramer, 2018 | Web of Science (all records) | All fields | 2010-2017 | Articles and reviews | 12.3 million | Unpaywall data integrated in Web of Science | 2017/12/20 – 2018/01/05 | Publication year, Subject areas, languages, countries, institutions, funders | ResearchGate not included. Preprints not included | Almost 30% OA for pub. year 2016 |

## 1.5.1 Google Scholar as a source of OA evidence

Google Scholar has become one of the most widely used tools for researchers to search scientific information (Bosman & Kramer, 2016; Mussell & Croft, 2013; Nicholas et al., 2017; Van Noorden, 2014a). By automatically parsing the entire academic web instead of indexing only some specific sources, Google Scholar's coverage is much more extensive than the coverage of any other multidisciplinary commercial databases like Web of Science and Scopus. Although there are not official figures on the size of its document base, it was estimated in approximately 170 million records in 2014 (Orduna-Malea, Ayllón, Martín-Martín, & Delgado López-Cózar, 2015). Recently, Google Scholar's chief engineer, Anurag Acharya, has declared that the size of its document base is "larger than the estimates that are out there" (Rogers, 2017).

An important feature of Google Scholar is that it usually provides links to freely available versions of the documents displayed in its results page, also when the document is not openly accessible from the publisher website. Unfortunately, despite the wealth of information available in Google Scholar, the platform does not provide a way to easily extract and analyse its data (something like an open API), reportedly because the agreements that Google Scholar had to reach with publishers to access their content preclude this (Van Noorden, 2014b). Perhaps because of this limitation, all OA-related studies based on Google Scholar data either used very small samples of documents, mostly focusing on specific case studies, or the samples of documents they analyzed were not random because the selection of documents relied on searches in the platform, and Google Scholar is known to rank documents primarily, although not only, on descending order of citations (Martin-Martin, Orduna-Malea, Harzing, & Delgado López-Cózar, 2017). Moreover, most of these studies only analyzed the links to freely accessible full texts that are displayed beside the primary version of the document in Google Scholar, but not the links available in the secondary versions (see Figure 2). Table 2 contains information on the sample of documents analyzed, source of OA evidence used, and OA levels found by studies that used Google Scholar as a source of OA evidence.

These studies all pointed to the value of Google Scholar as a source of free availability of scientific literature, but were limited in scope and thematically. Thus, it is still missing in the literature a relatively large-scale study of the free availability of scientific publications that can be identified through Google Scholar. This paper aims at filling this gap.

*Table 2. Studies that analyse OA levels using Google Scholar as a source of OA evidence*

| Study | Sample of documents | | | | | OA evidence | | | OA levels |
|---|---|---|---|---|---|---|---|---|---|
| | Source | Field | Pub. Year | Doc types | Size | Source | Date of data collection | Levels of aggregation | |
| Christianson, 2007 | Journals in CSA's Ecology Abs. and JCR: Ecology (random) | Ecology | 1945-2005 | Articles | 840 | Google Scholar | 2005/03 | Only total figure | 9% of the articles were freely accessible from Google Scholar |
| Norris, Oppenheim, & Rowland, 2008 | Web of Science (selected journals) | Ecology, Appl. Math., Sociology, Economics | 2003 | Articles | 4,633 | OAIster, OpenDOAR, Google, Google Scholar | Not declared | Subject area | Economics: 65%; Appl. Math.: 59%; Ecology: 53%; Sociology: 21%. Overall OA: 49% |
| Pitol & De Groote, 2014 | Web of Science (organization search) | Psychology, Chemistry, Electrical Engineering, Earth Sciences | 2006-2011 | Articles | 982 | Google Scholar | Not declared | OA version provider, OA type | 70% of documents were freely accessible in some form |
| Khabsa & Giles, 2014 | Microsoft Academic Search (random sample) | All fields | All years | Not specified | 1,500 (100x15) | Google Scholar | 2013/01 | Subject areas | Top OA categories: Computer Science (50%), Multidisciplinary (43%), Economics & Business (42%). Overall OA: 24% |
| Jamali & Nabavi, 2015 | Google Scholar (topic search) | All fields | 2004-2014 | All except citations and patents | 8,310 | Google Scholar | 2014/04 | Subject areas, OA types | Top OA category: Life Sciences (66.9%). Lowest OA category: Health Sciences (59.7%). Overall OA: 57.3% |
| Laakso & Lindman, 2016 | Scopus (selected journals) | Information Systems | 2010-2014 | Articles | 1,515 | Google, Google Scholar | 2015/02 | Journal, OA types | 60% of the articles were freely accessible from Google Scholar |
| Martín-Martín et al., 2016 | Google Scholar (pub. year search) | All fields | 1950-2013 | All types | 64,000 | Google Scholar | 2014/05 | Publication year | 40% of documents were freely accessible for the whole period. Over 66% considering only pub. years 2000-2009 |
| Teplitzky, 2017 | Pangaea (topic search) | Earth Sciences | 2010, 2015 | All types | 744+482 = 1,226 | Google Scholar | 2016/05 | OA types | 75% of documents in pub. year 2010, and 72% in pub. year 2015 |
| Abad-García, González-Teruel, & González-Llinares, 2018 | Web of Science (funding search) | Health | 2012-2014 | Articles | 762 | OpenAIRE, BASE, Recolecta, Google Scholar | Not declared | Only total figures | 46.3% of the documents were freely available from some source. Recall of Google was 93.5% |
| Mikki, Ruwehy, Gjesdal, & Zygmuntowska, 2018 | Web of Science (topic search) | Climate and ancient societies | All years | All types | 639 | Google Scholar | Not declared | Publication years | 74% of the documents were freely accessible |
| Laakso & Polonioli, 2018 | Publication lists of ethics researchers | Ethics | 2010-2015 | Articles | 1,682 | Google Scholar | 2017 | Publication years, OA types | 56% of the documents were freely accessible |

## 1.6. Research questions

This paper mainly intends to ascertain the suitability of the data available in Google Scholar to gauge the levels of adoption of OA in scientific journal articles, across all subject categories and countries, thus overcoming the limitations related to sample selection and sample size of the previous OA-related studies that used this source of data. Specifically, this article aims to answer the following questions:

- RQ1. How much of the recently published scientific literature is freely available according to the data available in Google Scholar, by year of publication, subject categories, and country of affiliation of the authors?
- RQ2. How much is openly accessible in a sustainable and legal way, and what proportion is freely available but does not meet these criteria?
- RQ3. What is the distribution of freely available documents by web domains?

# 2. Methods

The three main citation indexes of the Web of Science Core Collection (Science Citation Index Expanded [SCIE], Social Sciences Citation Index [SSCI], and Arts & Humanities Citations Index [A&HCI]) were used to select the sample of documents analysed in this study. All documents with a DOI indexed in either the SCIE, SSCI, or the A&HCI, and published in 2009 or 2014 were selected on the 19$^{th}$ of May, 2016. The rationale behind choosing these two years was that we wanted to analyse a large sample of documents from various publication years, but we also wanted to keep the sample manageable because of the difficulty of extracting data from Google Scholar. At the time of data collection, 2014 was the most recent year in which most articles scheduled to become OA after an embargo (Delayed OA) had already become OA. The data from articles published in 2009 would give us information on the trend.

The records of these documents were extracted from the local version of the Web of Science database available at the Centre for Science and Technology Studies (CWTS) in Leiden University. A total of 2,610,305 records were extracted, 1,080,199 from 2009, and 1,530,106 from 2014. We decided to use this source (as opposed to the CrossRef registry) because it would later enable us to carry out detailed analyses of the data, with breakdowns by subject categories, country affiliations, publication years, and journals.

It is worth noting that the number of Web of Science documents in these two years (2009 and 2014) at the time of writing this article had increased from 2,610,305 to 2,893,175. This could have been caused by backwards indexing of new documents, or by the addition of DOIs to records that previously did not contain one in the Web of Science database.

Each of these documents was searched on Google Scholar, using a non-documented method to search documents by their DOI. Example of query for the document with DOI "10.1010/j.jmmm.2013.09.059":

> https://scholar.google.com/scholar_lookup?doi=10.1010/j.jmmm.2013.09.059

Given that Google Scholar does not provide an API to query its database, a custom Python script was developed to carry out a query for each of the DOIs in our sample and scrape the data from the results page. Queries were distributed across a pool of different IP addresses to minimise the amount of CAPTCHAs (Completely Automated Public Turing test to tell Computers and Humans Apart) that Google Scholar requests users to solve from time to time. However, this approach did not entirely suppress the appearance of CAPTCHAs, which were solved manually when the system requested them. Additionally, when it was detected that Google Scholar provided a link to a freely accessible full text of a document, the link to the secondary versions of the same document was also followed through, in order to extract all the additional links to freely accessible full texts of the document that Google Scholar might have found (Figure 2). Searches were carried out off-campus to avoid retrieving links to full texts that are only accessible through library subscriptions. The process of extracting the data from Google Scholar was very time-consuming, taking over three months (from the end of May to the end of August of 2016) to collect data for the 2,610,305 selected documents.

*Figure 2. Example of primary and secondary versions of an article in Google Scholar*

Using the search strategy described above, Google Scholar retrieved results for 99.3% of the documents searched. The system did not retrieve any results for 0.7% of the DOIs searched. However, this does not necessarily mean that these documents were not covered by Google Scholar. These documents might have been covered by Google Scholar without a DOI, and therefore they might have been found using other search strategies, for example, searching by the title of the document. However, we did not try other search strategies, as we considered the results could not be overly affected by these missing documents.

A test was also carried out to assess the accuracy of the results retrieved from Google Scholar. That is, whether or not we had actually retrieved data about the documents we were looking for. In order to do this, we compared the bibliographic information available from Web of Science, with the data extracted from Google Scholar. The match was considered successful if at least one of the following criteria were met:
- Similarity of document titles in the two sources of data (based on the Levenshtein distance of the two strings of text) was equal or greater than 0.8 (similarity is 1 when the titles are exactly the same, and 0 when they are completely different).
- Similarity of document titles was between 0.6 and 0.8 AND the documents shared the same first author AND the same publication year.
- Same first author and same publication year, and title of document in Google Scholar was not in English. In some cases when the journal publishes in a language other than English, the title provided by Google Scholar is the original title, whereas in Web of Science, the title of the document is always displayed in English (even when the document itself is not written in English). In these cases the title similarity was very low, and using it resulted in a significant number of false negatives.

Based on these criteria, we classified as good matches 96% of the documents in our sample (2.51 million documents). The proportion of good matches was slightly higher if we only considered documents of the type "article" or "review" (97.6%). Therefore, we decided to analyze only the articles and reviews in our sample that we had considered as good matches, a total of 2,269,022 documents.

Google Scholar does not provide any information on the type of source that is providing free access to the full text of a document. For this reason, we combined information from a variety of sources in order to provide more detailed information about the type of free access that Google Scholar had been able to detect. We classified each *full text link* in one of the following categories:
- Publisher: when the full text is hosted on a publisher website, or on journal aggregators such as JSTOR or SciELO. Data from the [oaDOI dataset from 18 August 2017](#), DOAJ (Directory of Open Access Journals), and the Ulrich's Directory of Journals was used to create a list of websites where journal publishers make their articles available.
- Repository: when the full text is hosted in a repository, as defined by the Registry of Open Access Repositories (ROAR), and the Directory of Open Access Repositories (openDOAR).
- Research Institutions: when the full text is hosted in the web domain of a research institution (universities, research centers, institutes), excluding the website of the institutional repository. That is to say, this category mostly contains personal websites of individual researchers, research groups, departments, etc. inside an academic domain. In order to determine which domains belonged to academic institutions, a list of academic domains was also extracted from openDOAR.
- Academic Social Networks: in this category we only classified the full texts available from ResearchGate and Academia.edu.
- Harvesters: websites that copy full texts from other sources and make them available from their own servers. In this category we classified full texts hosted in the search engines CiteSeerX and Semantic Scholar, and the British CORE service.
- Non-categorized: any website that could not be classified in the previous categories.

After combining the information from the sources described above, there were still thousands of web domains that had not been classified. Therefore, we decided to manually check the hosts with a higher number of occurrences in our sample that still had not been categorised. Specifically, we checked the domains in which Google Scholar had found 100 or more full texts of documents in our sample, and the hosts that Google Scholar more frequently selected as the primary full text version (because these hosts would likely be publishers, as declared in Google Scholar's publisher guidelines[22]). Thus, approximately 1,000 hosts were classified after visiting the website and checking it manually. The rest of the web domains that had not been classified were considered as "non-categorized". The specific categorisation of hosts used in this study is available in the complementary material to this article[23].

In this article we make a distinction between Freely Available (FA) documents, and OA documents. We consider that all documents for which Google Scholar provided a link to a FA version of the document, regardless of the legality under which they were shared and their sustainability over time, are FA. When FA documents meet certain additional criteria (described below) they were also considered OA.

Unfortunately, there is no clear consensus regarding the minimum rights that any user should have in order to be able to consider a document OA. Some definitions, like the one declared by the BOAI or the Open Definition[24] are clear in that mere right to access the document free of charge is not enough to consider a document OA. They consider it necessary that the license extends other rights to all users, like redistribution, modification, or application for any lawful purpose. The reality, however, is that in many cases documents are made FA under licenses that fail to meet one or several of these criteria. For example, there are Creative Commons licenses that include Non-Commercial and/or Non-derivatives clauses, thus limiting the ways in which a document can be reused. The Elsevier user license[25] (the license under which Elsevier makes FA after an embargo period articles published in journals included in its Open Archive[26]) prohibits redistribution of the documents and reuse for commercial purposes. Moreover, there is a large portion of articles that publishers make available free of charge, without extending any other rights to users other than access. This is usually called "public access" in the publishing industry (Crotty, 2017). These issues have led some researchers to think in terms of degrees of openness, instead of considering OA a binary quality (Chen & Olijhoek, 2016).

Apart from the conceptual issues, there are also practical limitations for classifying documents as OA. In many cases, especially when we are talking about Green OA, there is no license attached to the document, or it is attached in a way that cannot be easily detected by automated systems. Fortunately, publishers are increasingly taking to sending license information to CrossRef (which makes these data openly accessible) or they display it as metadata in their own websites.

---

[22] https://scholar.google.com/intl/en/scholar/publishers.html#policies
[23] https://osf.io/fsujy/
[24] http://opendefinition.org/
[25] https://www.elsevier.com/open-access/userlicense/1.0
[26] https://www.elsevier.com/about/open-science/open-access/open-archive

For the reasons described above, in this article we use a more inclusive definition of OA than the one declared by the BOAI or the Open Definition, and we instead set our focus on sustainability and legality. Specifically, this article considers the following types of OA:

- <u>Gold OA</u>: when the journal that published the article was listed in DOAJ.
- <u>Hybrid OA</u>: when the journal was not listed in DOAJ but an OA license was recorded in the metadata available in CrossRef, and the Open license came into effect at the same time the article was published (OA immediately upon publication). We considered as OA licenses all Creative Commons licenses, the Elsevier OA user license, and other OA licenses registered in CrossRef by publishers like the ASPB[27], ACS[28], and IEEE[29]. Our operational definition of "OA immediately upon publication" was that the value recorded in the *delay-in-days* field of the License element available in the CrossRef metadata (defined as the "[n]umber of days between the publication date of the work and the start date of this license"[30]), should be less than 30 (one month). We decided to set this limit instead of *delay-in-days* = 0 because we noticed that for some articles published as OA, the Open license came into effect a few days after publication, and we considered that these articles should also be classified as "OA immediately upon publication".
- <u>Delayed OA</u>: when the journal was not listed in DOAJ but an Open Access license was recorded in the metadata available in CrossRef, and the Open license came into effect more than 30 days after the publication of the article.
- <u>Bronze OA</u>: when the full text is FA from the publisher, but the journal is not listed in DOAJ and no OA license could be found. This category includes gratis / public access from the publisher (free to read but the publisher retains copyright), but might also contain masked Hybrid or Delayed OA (when the publishers fail to disclose an OA license in machine-readable form), and possibly even some masked Gold OA (if a full OA journal is not listed in DOAJ and the publisher does not discloses an OA license).
- <u>Green OA</u>: the documents that are FA from institutional or subject-based repositories, as listed in ROAR and OpenDOAR.

All the documents that were available from sources other than the publisher website and repositories (such as websites of research institutions excluding the repository, academic social networks, harvesters, and the rest) were only considered as FA, and not OA. We took this conservative measure because we wanted to make a distinction between more legally sound and sustainable sources (publishers and repositories) which are more likely to be copyright-compliant and usually implement long-term preservation plans for the documents they host, and less stable sources (personal websites, academic social networks…) where any document (regardless of its copyright status) can be uploaded and deleted at any time.

Lastly, Google Scholar does not provide data on the publication stage of the freely accessible versions that it finds: that is, whether the free version is a preprint (before peer-review), an author's accepted manuscript (after peer-review, but before typesetting), or the journal's version of record (final published article). Although this is an interesting aspect of OA publishing, identifying the type of version would have required accessing the full text of each individual article, and so it falls outside the scope of this study.

---

[27] https://aspb.org
[28] https://pubs.acs.org
[29] https://www.ieee.org
[30] https://github.com/CrossRef/rest-api-doc/blob/master/api_format.md

Data was processed and analyzed using the R programming language. The percentages of OA documents were computed by publication year, subject category, country of affiliation (considering all co-authors), and journal. The data used in this study is openly available (Martín-Martín, Costas, van Leeuwen, & Delgado López-Cózar, 2018). This will facilitate the creation of custom analyses that focus on the research done in specific countries, specific fields, specific journals, etc.

# 3. Results

## 3.1. General overview

Google Scholar provided links to FA full texts for 54.7% of our sample of documents (Figure 3). If we break down the results by year of publication, documents published in 2014 show a slightly higher percentage of FA documents (55.8%) than documents published in 2009 (53%), even though the number of documents published in 2014 (1,331,795) was larger than the number of documents published in 2009 (937,227), and the fact that at the time of data collection documents from 2014 had had considerably less time to be made freely available on the Web than documents from 2009.

*Figure 3. Overall OA and FA levels found in Google Scholar, by year of publication and both years combined*

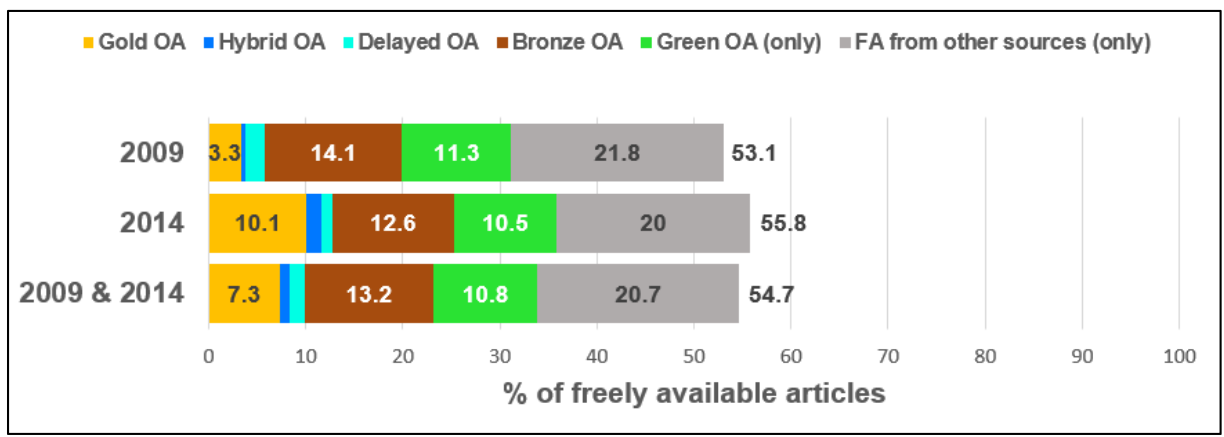

If we consider the two years under study (Figure 3), we can see that 23.1% of the documents are FA from publisher websites (Gold + Hybrid + Delayed + Bronze). It is worth noting that most of the documents available from publishers are Bronze OA, which are usually made accessible under very restrictive reuse terms. However, it seems like Gold and Hybrid are gaining importance, judging by the increment from 3.3% to 10.1% from 2009 to 2014 of Gold OA, and from 0.5% to 1.5% for Hybrid OA. Bronze OA decreased from 14.1% to 12.6%, and Delayed OA decreased as well (from 2% in 2009 to 1.1% in 2014).

Figure 3 displays OA provided by the publisher (Gold, Hybrid, Delayed, Bronze), Green OA, and FA from other sources. However, in the cases where a document is available from several types of sources, publisher-provided OA is given preference over Green OA and FA from other sources. In a similar manner, Green OA versions are given preference over FA from other sources. Therefore, Figure 3 does not display the total percentages of Green OA and FA from other sources. These are displayed in Figure 4.

The proportion of documents available as Green OA (repositories) was higher in the publication year 2014 (18.9%) than in 2009 (15.7%), as displayed in Figure 4. However, the number of documents that were available from repositories and not from the publisher (displayed in Figure 3) was slightly higher in the publication year 2009 (11.3%) than in 2014 (10.5%).

*Figure 4. Total percentage of Green OA and FA found in Google Scholar, by year of publication and both years combined*

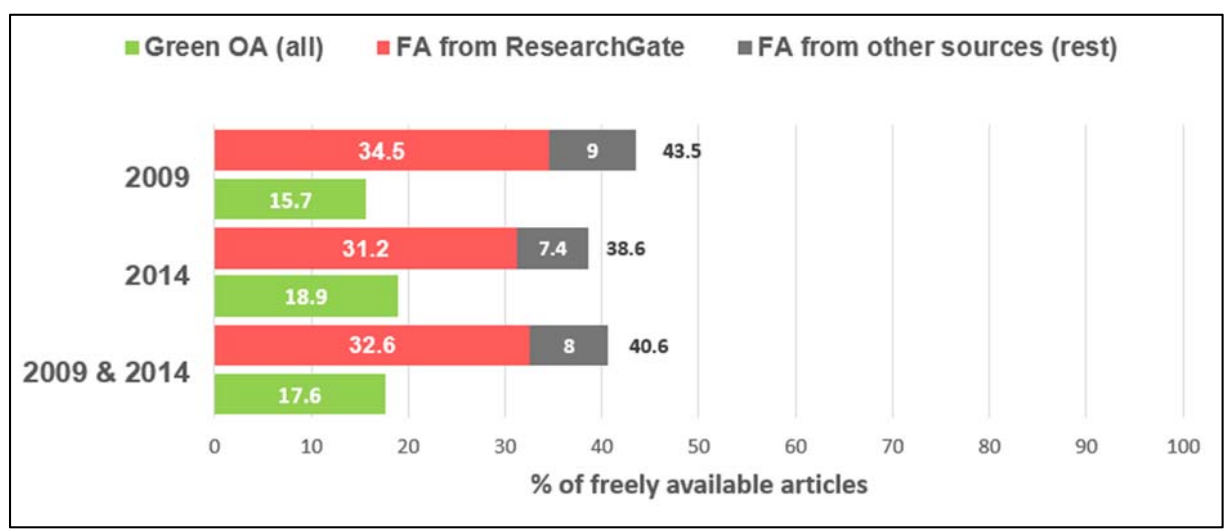

Apart from publisher websites and repositories, there is a large fraction of documents that are available from other sources (mainly the academic social network ResearchGate, but also personal websites, and harvesters). Google Scholar found that 43.5% of the documents in the sample published in 2009 were available from other sources (Figure 4). This percentage was lower in the publication year 2014 (38.6%). Nevertheless, in both years this percentage is larger than the sum of what all publishers and repositories together provided. Moreover, a considerable portion of these documents are FA only from these other sources (that is, these documents are not openly accessible from the publisher or from repositories). This figure remains relatively stable in the two publication years (21.8% in 2009, and 20% in 2014), as can be observed in Figure 3.

The predominance of sources other than publishers and repositories can also be observed if we take a look at the number of freely available documents by website (Table 3). By far, the source that provided more freely available full texts was the academic social network ResearchGate, which by itself provided access to 32.6% of the documents in our sample (738,573). If we compare this figure to the percentage of documents provided as OA by publishers available in Figure 3 (23.1%, approx. 525,000 documents), we see that ResearchGate provided access to more documents in our sample than all publishers together. Moreover, 32.7% of the documents available from ResearchGate (over 240,000) were not freely available from any other source.

Table 3 also shows how often Google Scholar displays links from each host as the primary full text links. This is interesting because the primary link is likely to be the link that most Google Scholar users click to access the full text of an article. Again, ResearchGate is first in the rank, followed by Pubmed Central (www.ncbi.nlm.nih.gov) and arXiv. However it is worth noting that some hosts that provided many FA documents (europepmc.org, academia.edu, citeseerx.ist.psu.edu) are rarely selected by Google Scholar as the primary full text links (only in 10.3%, 14.1%, and 9.3% of the cases, respectively), meaning that the documents these platforms provide are also available from other platforms which are placed higher in Google Scholar's host precedence rules. Regarding these precedence rules, the data in Table 3 shows that Google Scholar does indeed tend to select the publisher version as the primary version whenever it is an option (as stated in its indexing policies). Most publisher websites are selected as the primary full text version in over 90% of the cases. The exceptions seem to be Springer and BioMed Central, which are only selected as the primary version in about 45% of the cases. Lastly, it appears that Google Scholar chooses the arXiv repository even over most publishers, as this repository is selected as the primary source of full text in 99.9% of the cases. This means that when an article is openly accessible from arXiv, Google Scholar always chooses the arXiv version as the primary full text version, presumably even when the article is also openly accessible from the publisher.

*Table 3. Top 20 websites according to the number of FA full texts they host.*

| Host | Type | # of FA documents | % as only FA provider | # of FA as primary version | % as primary version |
|---|---|---|---|---|---|
| www.researchgate.net | Social network | 738,573 | 32.7 | 323,372 | 43.8 |
| europepmc.org | Repository | 177,930 | 5.1 | 18,312 | 10.3 |
| www.academia.edu | Social network | 168,485 | 4.2 | 23,681 | 14.1 |
| www.ncbi.nlm.nih.gov | Repository | 165,403 | 1.8 | 74,109 | 44.8 |
| citeseerx.ist.psu.edu | Harvester | 120,378 | 1.8 | 11,203 | 9.3 |
| arxiv.org | Repository | 72,862 | 25.0 | 72,753 | 99.9 |
| onlinelibrary.wiley.com | Publisher | 49,887 | 32.8 | 47,712 | 95.6 |
| www.sciencedirect.com | Publisher | 47,356 | 26.1 | 43,825 | 92.5 |
| pdfs.semanticscholar.org | Harvester | 38,164 | 1.0 | 2,790 | 7.3 |
| journals.plos.org | Publisher | 37,984 | 12.5 | 37,380 | 98.4 |
| link.springer.com | Publisher | 35,295 | 6.2 | 15,335 | 43.4 |
| www.biomedcentral.com | Publisher | 27,400 | 2.1 | 12,328 | 45.0 |
| www.nature.com | Publisher | 23,726 | 26.1 | 21,699 | 91.5 |
| downloads.hindawi.com | Publisher | 18,566 | 38.8 | 18,565 | 100.0 |
| core.ac.uk | Harvester | 15,344 | 1.4 | 769 | 5.0 |
| pubmedcentralcanada.ca | Repository | 14,286 | 1.0 | 461 | 3.2 |
| hal.archives-ouvertes.fr | Repository | 11,293 | 10.7 | 5,530 | 49.0 |
| www.mdpi.com | Publisher | 11,084 | 12.9 | 11,083 | 100.0 |
| www.infona.pl | Repository | 10,060 | 41.4 | 6,132 | 61.0 |
| www.tandfonline.com | Publisher | 8,973 | 61.2 | 8,730 | 97.3 |

## 3.2. Analysis by disciplines

We mapped the original WoS subject categories to more general classification schemes: one containing 7 broad subject areas, and the other containing 35 scientific disciplines. The schemes were introduced by Tijssen et al. (2010), and the specific correspondence with WoS categories is available in the complementary materials.

There is a high inter-area variability, ranging from 60% overall availability in the Medical and Life Sciences, to 32.3% overall availability in Law, Arts, and Humanities (Figure 5). Multidisciplinary journals achieve a 93.6% overall availability, which is natural if we consider that this category includes Gold OA multidisciplinary mega-journals such as PLOS ONE.

*Figure 5. OA and Free Availability levels found in Google Scholar, by broad subject areas.*

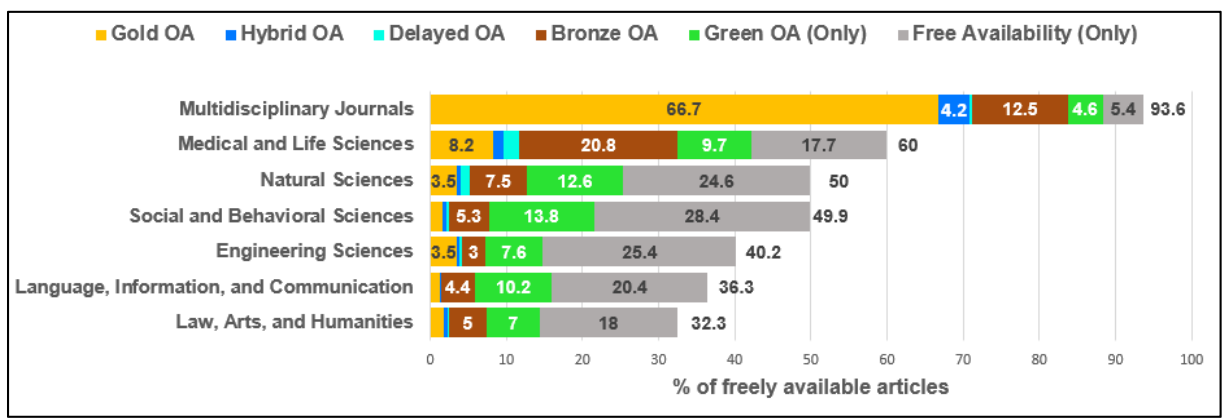

If we descend to the level of disciplines (Figure 6) we can see that Bronze OA is usually the predominant type in which publishers provide OA. In 28 out of the 35 disciplines shown in Figure 6, the percentage of Bronze OA is higher than the sum of Gold, Hybrid, and Delayed OA. Bronze OA is especially important in Basic Life Sciences, Biomedical Sciences, and Clinical Medicine.

*Figure 6. OA and Free Availability levels found in Google Scholar, by scientific discipline*

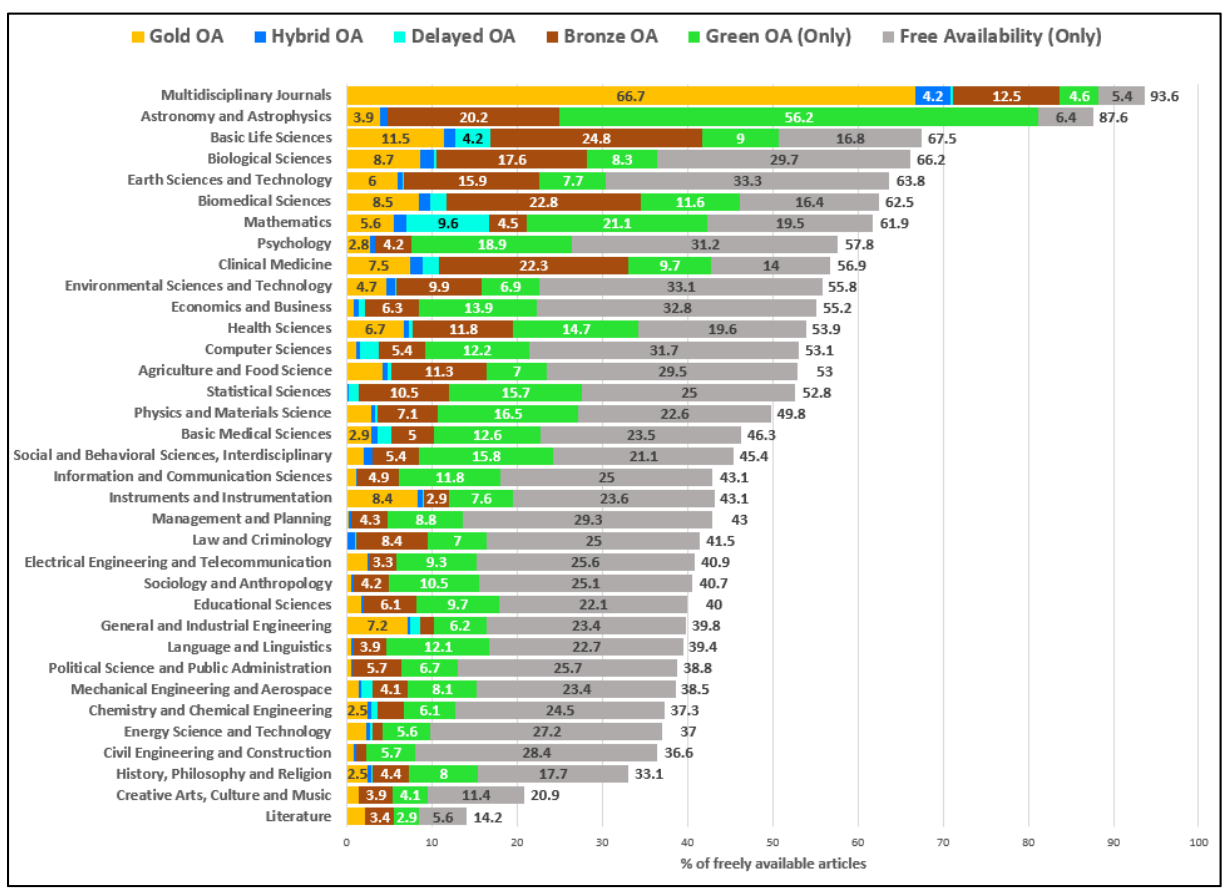

Figure 6 also shows the percentage of articles in Green OA that are not openly accessible from the publisher: Green OA (only)[31]. In 19 out of the 35 disciplines, the number of documents that are accessible only through Green OA was higher than the sum of Gold, Hybrid, Delayed, and Bronze OA. The disciplines with a larger share of documents in the Green OA (only) category are Astronomy and Astrophysics (56.2%), and Mathematics (21.1%).

If we consider FA only (the cases when documents were only available from sources other than publishers and repositories), Figure 6 shows that this is the most frequent type of availability in most disciplines. In 23 out of the 35 disciplines, FA (only) achieves higher percentages than Gold, Hybrid, Delayed, Bronze, and Green combined. In four of these disciplines (Management and Planning, Political Science and Public Administration, Energy Science and Technology, and Civil Engineering and Construction), more than two thirds of the documents that were FA in some form, were only available from sources other than the publisher or repositories.

Lastly, it is worth noting that there is a large degree of intra-discipline variability as well. Figure A2 in the complementary materials[32] displays the correspondence between the 35 disciplines in Figure 6, and the subject categories used by the Web of Science. This figure shows that in many cases there are important differences among the categories of a discipline, regarding not only the overall free availability of documents, but also the types of availability. If we take Clinical Medicine (56.9% overall free availability), for example, the subject categories with the highest overall availability are *Tropical Medicine* (85.9%), and *Andrology* (84.7%). Both categories also present high levels of OA provided by the publisher (over 70%). *Dermatology,* however, presents a completely different behavior: only 37% of the documents are freely available in some way, and the most common type of availability is FA from other sources (14.5%).

---

[31] Total percentages of Green OA by subject categories (including the cases when the article is also openly accessible from the publisher) are available from the complementary materials and in the web application.

[32] https://osf.io/fsujy/

## 3.3. Analysis by countries of affiliation

Table 4 displays OA and FA levels of countries with an output equal or higher than 1% of the total, considering only documents published in 2014 (the most recent year in our sample). The affiliation of all co-authors of the articles were considered (each article was considered once for each different country of affiliation). It distinguishes between OA provided by the publisher, OA from repositories (when OA from publisher is not available), and FA from any other sources (when OA from publisher or from repositories is not available). A green background in one of the cells of the table indicates that the value in that cell is higher than the World value (visible in the first row below the headers). A red background indicates a value lower than the World value. Higher color intensity indicates a higher distance relative to the World value. The last column (% OA + FA) highlights the top three countries with a higher overall availability (in green) and the top three countries with a lower overall availability (in red).

*Table 4. OA and Free Availability (FA) levels for documents published in 2014 by researchers in countries with high output (>1% of the total)*

| Country | Documents | % OA from publisher | % OA from repositories* | % OA Total | % FA other sources† | % OA + FA† |
|---|---|---|---|---|---|---|
| **World** | 1,331,795 | 25.3 | 10.5 | 35.8 | 20.0 | 55.7 |
| **USA** | 360,889 | 29.1 | 18.2 | 47.3 | 18.9 | 66.2 |
| **Peoples R China** | 231,162 | 22.9 | 4.3 | 27.2 | 18.7 | **46.0** |
| **Germany** | 96,265 | 28.6 | 13.4 | 42.0 | 19.2 | 61.3 |
| **England** | 89,996 | 35.0 | 15.9 | 50.9 | 17.3 | 68.3 |
| **Japan** | 71,587 | 26.6 | 9.9 | 36.5 | 13.4 | 49.9 |
| **France** | 66,648 | 26.5 | 17.4 | 43.9 | 23.5 | 67.4 |
| **Canada** | 60,342 | 28.1 | 10.5 | 38.6 | 23.1 | 61.7 |
| **Italy** | 58,397 | 26.2 | 11.9 | 38.1 | 25.6 | 63.7 |
| **Australia** | 53,822 | 26.2 | 10.5 | 36.7 | 24.9 | 61.7 |
| **Spain** | 51,586 | 25.3 | 13.9 | 39.2 | 24.7 | 63.9 |
| **South Korea** | 51,036 | 26.2 | 5.4 | 31.6 | 17.9 | 49.5 |
| **India** | 50,468 | 15.7 | 7.4 | 23.1 | 25.6 | 48.7 |
| **Netherlands** | 36,228 | 33.7 | 14.2 | 47.9 | 22.9 | **70.8** |
| **Brazil** | 34,517 | 37.0 | 8.8 | 45.8 | 25.8 | **71.6** |
| **Russia** | 28,108 | 10.6 | 9.7 | 20.3 | 23.9 | **44.3** |
| **Switzerland** | 26,580 | 33.8 | 14.9 | 48.7 | 21.8 | **70.5** |
| **Taiwan** | 25,492 | 27.3 | 8.4 | 35.7 | 17.5 | 53.2 |
| **Sweden** | 24,286 | 35.3 | 14.9 | 50.2 | 19.2 | 69.4 |
| **Iran** | 23,387 | 14.5 | 4.1 | 18.6 | 26.4 | **45.0** |
| **Turkey** | 21,516 | 22.8 | 5.8 | 28.6 | 23.9 | 52.5 |
| **Poland** | 20,496 | 33.4 | 9.6 | 43.0 | 20.7 | 63.8 |
| **Belgium** | 19,809 | 29.5 | 15.7 | 45.2 | 24.2 | 69.4 |
| **Denmark** | 15,853 | 34.9 | 12.4 | 47.3 | 20.2 | 67.5 |
| **Scotland** | 13,813 | 38.3 | 18.3 | 56.6 | 16.4 | <u>73.0</u> |
| **Austria** | 13,514 | 34.9 | 12.2 | 47.1 | 19.3 | 66.4 |

\* Accessible from repository but not from publisher
† Only available from other sources

All countries in Table 4 present higher percentages of OA from publishers than of OA only from repositories. 18 out of the 25 high output countries displayed in Table 4 present OA levels (sum of OA from publisher and OA from repositories) that are higher than the World level (35.8%). 13 of these countries are in Europe. The other five are the USA, Japan, Canada, Australia, and Brazil. The countries with the highest percentages of OA come very close to or slightly surpass 50% of the total amount of documents published by researchers in that country (United Kingdom, Sweden, Switzerland, Netherlands, USA, Denmark, Austria). All these countries present percentages of OA from publishers and from repositories that are higher than the average world percentages. Japan, Brazil, and Poland also have higher than average OA levels, with the particularity that most of their OA is available from publishers, and their percentage of OA from repositories is lower than the World level. The opposite, however, does not occur: there are no countries in Table 4 with a lower than average percentage of OA from publishers that manage to achieve a higher than average total percentage of OA thanks to OA from repositories.

7 out of the 25 high output countries displayed in Table 4 present OA levels that are lower than the World level (35.8%). Chief among them is China, with only 27.2% of its documents accessible either from the publisher or from repositories, even though it is the second country in terms of output (231,162 articles and reviews published in 2014). The other six countries are also located in Asia (South Korea, India, Russia, Taiwan, Iran, and Turkey).

At the world level, 20% of the documents are only freely available through sources other than publishers and repositories. At the country level there is some variation: from the 13.4% percent of documents written by Japanese researchers that are only available from these other sources, to the cases of Italy, India, Brazil, and Iran, where the percentage is slightly over 25%.

If we consider overall availability (the sum of OA and FA only), the countries with a higher percentage of availability are Brazil (71.6%), the Netherlands (70.8%), and Switzerland (70.5%). Scotland deserves a special mention, because if considered separately from the rest of the United Kingdom (which is the way the Web of Science presents authors' affiliations), it achieves 73% overall free availability. The United Kingdom as a whole presents a slightly lower percentage (68.7%). In the lowest positions of the rank we can find China (46%), Iran (45%), and Russia (44.3%).

Table A1, available in the complementary materials[33], extends Table 4, displaying the same information for 40 additional countries, those with an output larger than 0.1% and lower than 1% of the World total. The countries with a higher overall availability in this output tier are Kenya (1,504 documents, 80.6% overall availability), Chile (5,812 documents, 76% overall availability), and Norway (11,601 documents, 67.9% overall availability), and the countries with a lower overall availability are Tunisia (3,008 documents, 50.3% overall availability), Ukraine (4,397 documents, 49.1% overall availability), and Algeria (2,139 documents, 43.1% overall availability).

# 4. Discussion

## 4.1. Limitations and further lines of study

The analysis carried out in this study suffers from a number of limitations. These are related either to the sample selection, to the data available in Google Scholar, to the categorisation of OA / FA of the documents in the sample, or to the replicability of the study.

The first limitation of this article related to sample selection is that it only analyses scientific journal articles and reviews published in journals indexed in Clarivate Analytics' SCIE, SSCI, and A&HCI. These three citation indexes are known to have limited coverage of journals in the Social Sciences, Arts, and Humanities (SSAH), and to suffer from a bias towards English-language journals (Mongeon & Paul-Hus, 2016; Van Leeuwen, Moed, Tijssen, Visser, & Van Raan, 2001). Therefore, results might have been different if more articles published in journals in the SSAH that are not covered by these indexes, and/or more articles from journals that publish in languages other than English had been included in the sample. Furthermore, this study focuses on the OA levels of articles and reviews, and not on the OA levels of other document types such as books, conference papers, or scientific reports. Further studies could focus on the free availability of these other document typologies, which Google Scholar also covers.

---

[33] https://osf.io/fsujy/

An additional limitation is that this article only considers articles and reviews for which a DOI was available in Clarivate Analytics' citation indexes at the time of data collection. Documents without a DOI, or documents for which a DOI had been minted but was not recorded in these databases at the time of data collection, have not been considered in this study.

Regarding the data extracted from in Google Scholar, this study has the following limitations:
1. This study only analyses OA evidence in Google Scholar of documents published in 2009 and 2014 at a specific moment in time: summer of 2016. Therefore, no extrapolation should be made regarding OA levels of other publication years. Furthermore, OA levels of documents published in 2009 and 2014 might have changed by the time of this writing, caused by OA backfilling: documents that have become OA after we collected the data, either because the publisher practices Delayed OA, or because authors have self-archived their articles. It also may be the case that some documents that were available when we collected the data are no longer available. The dispute between the Coalition for Responsible Sharing and ResearchGate, in which ResearchGate was forced to remove from public view a significant number of articles that infringed copyright, may have affected the current levels of free availability of the documents in our sample. Additionally, some documents hosted in other unstable sources, such as personal websites, may have also been removed.
2. In some cases, Google Scholar failed to recognize that an article was freely available from a source that the search engine indexes. In practice, this takes the form of a record in which no FA link is provided to the right of the main bibliographic information (see Figure 2), but if users would follow the link available in the title of the document, they would find that the article is in fact freely available. Our study only considers the links that Google Scholar provides to the right of the bibliographic information, and therefore, our results undercount free availability in these cases. We are aware that some journals (for example, some Gold OA journals published by *Frontiers*, and also *eLife*) were affected by this problem. We have also noticed that Google Scholar has fixed these errors for the most part, and at the time of this writing, FA links are correctly displayed to the right of the bibliographic information of the articles published in the aforementioned journals.
3. In some cases, Google Scholar is not able to successfully merge all the different versions of an article that can be found on the Web (Martín-Martín et al., 2014; Orduna-Malea, Martín-Martín, & Delgado López-Cózar, 2017), and as a result, two or more entries might exist in Google Scholar for documents that are actually the same. This might happen for a number of reasons, but is more frequent in journals that publish several versions of the same document (i.e. versions in several languages), and also for journals that, even though they publish only in one language, create versions of the article metadata in several languages. In these cases, Google Scholar's algorithms to detect duplicate documents usually fail. For our study this means that in some cases, the record we retrieved from Google Scholar might be one that does not provide a link to a freely available version, even though other entries of the same document in Google Scholar might contain such links. Therefore, our study undercounts free availability in these cases as well. One journal in our sample that is affected by this problem is *Revista Espanola de Documentacion Cientifica*, a Gold OA journal for which our data shows FA links in only 56% of the documents it published in 2009 and 2014.

Regarding the categorization of documents as OA / FA, and its specific subtypes (Gold OA, Hybrid OA, Delayed OA, Bronze OA, and Green OA, as well as FA only from sources other than the publisher and repositories), there are several limitations that should be taken into account.

1. We considered as Gold OA only the articles published in journals included in the Directory of Open Access Journals (DOAJ). There are, however, journals that adhere to the Gold OA model that are not included in this directory, like, for example, some journals owned by the Korean Association of Medical Journal Editors (*Korean Journal of Radiology*, and *Korean Journal of Physiology & Pharmacology*, for example). Because our study relies on DOAJ, it suffers from this limitation, and articles published in these journals are miscategorized either as Hybrid OA (when an Open License could be found) or as Bronze OA (if the journal does not deposit license information in CrossRef) Therefore, our study might be overestimating Hybrid and Bronze OA in detriment of Gold OA. Nevertheless, the error introduced by this issue in our calculation of Hybrid OA is estimated to be fairly small, as the total sum of articles in journals where more than 70% of the articles have been categorized as Hybrid OA (those that could be affected by this problem) is only 9,211 (0.4% of the sample).
2. The license information provided by CrossRef is incomplete. We found that for approximately 85,000 out of 163,000 articles classified as Gold OA (because the journal where they are published is listed in DOAJ), no open license was reported via CrossRef, suggesting that a large number of journals still do not deposit license information in CrossRef. If the proportion of Hybrid or Delayed OA journals that do not deposit license information in CrossRef is any similar, our results would be affected in that some Gold, Hybrid, and/or Delayed OA articles would have been erroneously classified as Bronze OA. Therefore, further analyses are needed to ascertain the specific composition of the Bronze OA category. It may turn out that Bronze OA is only a mix of Gratis Access provided by the publishers, and Gold, Hybrid or Delayed OA in journals that do not declare licenses in a easily identifiable way. In that case, the term "Bronze OA" will stop being necessary once these practical limitations are overcome.
3. Regarding Green OA, in this article we make the assumption that documents available from repositories are sustainable and legal. This might not be true in some cases, and therefore a more in-depth study of the sustainability and legality of subject and institutional repositories all over the world would be helpful to advance our knowledge of OA.
4. This study does not differentiate between the various versions of the articles that may have been made available on the Web: preprints that still have not gone through peer-review, authors' accepted manuscripts, and the publisher's version of record. Further studies are needed to detect the extent to which preprints are prevalent in specific subject areas, and whether this could affect the quality and validity of research that cites preprints, rather than accepted manuscripts or the publisher's version of record, which have been vetted by peer-review panels.

Lastly, perhaps one of the most important limitations of this study is that it is not easily replicable because of the limitations on data extraction imposed by Google Scholar. Extracting a large amount of data from this source is still only possible if one is willing to commit an inordinate amount of time to the task (three months, in our case). However, the goal of this study was not to describe a replicable method to analyze OA levels using Google Scholar, but to find out whether the data available in Google Scholar could in fact be useful for this purpose. If it turns out that the data *is* useful, a request could be made to Google Scholar to reconsider making their data (at least to the parts related to the free availability of documents) more open for reuse. Repositories have traditionally been in favor of interoperability (as proven by the OAI-PMH initiative), and publishers are slowly but steadily making article metadata more open through platforms like CrossRef and also thanks to initiatives like I4OC[34] (Initiative for Open Citations), so it is not clear who, if anyone, would be against opening these data nowadays. Of course, this would implicate a change of direction for a platform that has traditionally been quite reluctant to provide its data in bulk. It is possible that the Google Scholar team prefers to spend its efforts in the same problem they have been trying to solve up to now: connecting users with the academic documents they need to help them solve important problems. Nevertheless, as worthy as that goal is, it is also beyond doubt that these data would be of great interest to all actors in the scientific community, and might also be able to save duplicated efforts to other OA-related initiatives.

Despite the limitations described above, this study analyses the largest sample of data extracted from Google Scholar to date, and by combining these data with the data available in other sources such as DOAJ, CrossRef, OpenDOAR, and ROAR, it offers insights into all the variants of OA (Gold, Hybrid, Delayed, Bronze, Green). It also provides information on the free availability of documents from other sources (FA), thus providing a holistic, large-scale, and detailed depiction of the status of OA of scientific publications across all scientific fields and countries.

## 4.2. Comparison of results with similar studies

The report recently published by Science-Metrix (2018), and the studies published by Piwowar et al. (2018), Bosman & Kramer (2018), and van Leeuwen et al. (2017) are perhaps the ones that offer more opportunities for comparison with this study. This is because they all extracted samples of documents from the Web of Science. Moreover, they all analysed documents from 2009 and 2014 (among other publication years). In the case of Science-Metrix's report, they declare to have carried out their data collection in the third quarter of 2016, roughly the same months in which we carried out our own data collection. The two studies based on Unpaywall as well as van Leeuwen's study used data extracted more recently (2017), and thus differences between our study and theirs may be attributable at least in part to the backfilling that has occurred between the time of our data collection and theirs.

Science-Metrix reports 55% overall free availability both in 2009 and 2014. These results are very similar to ours (53.1% and 55.8% in 2009 and 2014, respectively). Their percentages on OA provided by the publisher are also very similar to ours: 20.2% in 2009, and 23.3% in 2014 in the Science-Metrix report, while our study shows 19.9% in 2009 and 25.3% in 2014 (Figure 3). The figures on Green OA differ in the two studies. The Science-Metrix report finds 33.3% and 31.5% of Green OA in 2009 and 2014, whereas our study only finds 15.7% and 18.9% in these years (Figure 4). The reason of this difference is probably that the Science-Metrix report considered documents available from ResearchGate as Green OA, and our study does not. However, our study shows that 34.5% and 31.2% of the documents in 2009 and 2014, respectively, are available from ResearchGate (which we label as FA only), which matches the results found by Science-Metrix.

---

[34] https://i4oc.org/

As regards the results at the country level, the country tables available in the Science-Metrix report offer strikingly similar results to the ones displayed in Table 4, although the percentages in our study are roughly 3 points higher for each country than in the Science-Metrix report (except in the case of Brazil, which has a higher percentage in the Science-Metrix report). The case of Brazil reveals other possible differences between Science-Metrix's approach, and ours, because they declare that SciELO is almost tied to ResearchGate in the number of freely accessible documents they offer, whereas our data shows that ResearchGate offers over 24,000 documents published by Brazilian researchers, and SciELO only 6,000.

As for the results at the level of subject areas, their results also agree with our study in that the areas with a higher percentage of free availability are the Health and Natural Sciences (over 50%), followed by Applied and the Social Sciences (between 40% and 50%), and lastly, the Arts & Humanities, with lower percentages (less than 40%).

Lastly, it is worth noting that the Science-Metrix study applies a calibration factor of 1.2 to the counts of freely available documents found by the 1science database, because the recall of this source is considered to be low. Therefore, although the results in this study closely match the results in the Science-Metrix study (at least at the levels of countries and broad subject categories), Google Scholar seems to have a better recall than the 1science database, because no calibration factor was applied in this case.

As stated in the literature review, the study by Piwowar et al. (2018) used three different samples, and one of them was a sample of documents covered by the Web of Science. Although their paper only reports the percentage of overall availability for documents in this sample (36.1%), the supplemental data they released alongside the paper (Piwowar et al., 2017) provides the necessary data to calculate OA percentages by year and type of OA in their WoS sample. Their data shows that 33.1% of the documents published in 2009, and 37.4% of the documents published in 2014 in their sample of WoS documents were freely accessible in some way according to Unpaywall. These results are very similar to ours (31,2% in 2009 and 35.8% in 2014), if we disregard the percentage of documents that we considered FA only (available only from sources other than publishers and repositories), which their study does not analyze. The slightly higher percentage in their study might be caused by slightly better coverage of OA sources in the Unpaywall system than in Google Scholar, but might also be explained by sampling issues (they use a sample, rather than the entire collection of documents), by small methodological differences regarding OA labelling (our study does not consider as Green OA documents hosted in personal or department websites inside academic domains, while theirs does), or by the fact that their study analyses data extracted in 2017, and therefore OA levels might have increased because of backfilling since the data in our study was collected (summer of 2016). In any case, the specific percentages of the different types of OA in the two studies are remarkably similar, as can be observed in Table 5.

*Table 5. Comparison of OA levels found by Google Scholar in this study, and by Piwowar et al. (2018) using Unpaywall data*

|                | 2009          |           | 2014          |           |
|----------------|---------------|-----------|---------------|-----------|
|                | Google Scholar | Unpaywall | Google Scholar | Unpaywall |
| **% Gold**     | 3.3           | 3.1       | 10.1          | 9.4       |
| **% Hybrid**   | 0.5           | 3.4       | 1.5           | 5.2       |
| **% Delayed**  | 2             | -         | 1.1           | -         |
| **% Bronze**   | 14.1          | 14.7      | 12.6          | 11.6      |
| **% Green (only)** | 11.3      | 11.9      | 10.5          | 11.2      |
| **% Total OA** | 31.2          | 33.1      | 35.8          | 37.4      |

Bosman & Kramer (2018) analysed the data from Unpaywall that the Web of Science has integrated into its system. They found an overall 28% of OA for documents published in 2014 (Kramer & Bosman, 2018). However, the Web of Science only provides OA information when the version that is FA is the author accepted manuscript (AAM), or the publisher's version of record (VOR). Therefore, this suggests that almost 10% of the documents covered by Unpaywall (and probably also Google Scholar, given the similarities found above) are preprints, that is, manuscripts that still have not gone through peer-review.

The results of this study show significantly higher percentages of OA (up to 15 points higher) than those found by van Leeuwen et al. (2017). That study reports overall OA levels of roughly 21% in 2009 and 27% in 2014. These differences may be explained by the more restricted approach of van Leeuwen's method, focused on OA sources related with the idea of legality and sustainability such as OpenAIRE, DOAJ, PubMed Central, etc., and with a strong focus on Gold and Green OA; while Google Scholar, Unpaywall and Science Metrix identify also Hybrid, Delayed and particularly Bronze OA. Considering together the Gold and Green OA shares in 2014 in this study (10.1%+10.5%) we come up with a closer value to the 27% observed in van Leeuwen's study, thus suggesting the relative consistency among methods, but also highlighting the role that Hybrid, Delayed and Bronze OA (together with FA only) play in the overall consideration of what is OA.

Lastly, the results from this study somewhat differ from those found by Jamali & Navabi (2015), who carried out a series of subject queries in Google Scholar to analyze OA levels in 277 minor subject categories extracted from Scopus. They found approximately 60% of free availability for documents published between 2004 and 2014 in all areas of research (Life, Physical, Social, and Health Sciences). This differs from our study, where we found significantly less free availability in the Social Sciences and Applied Sciences, than in the Natural and Health Sciences. The difference might be explained at least in part by the fact they only analyzed the first ten hits of each query, and Google Scholar is known to rank documents in a search based primarily on the number of citations that the documents have received (Martin-Martin et al., 2017). Highly cited documents might have different patterns of behavior regarding OA availability than a randomly selected sample of documents. Moreover, their study was not limited to documents covered by the Web of Science, which might also have influenced the results.

# 5. Conclusions

## 5.1. Answers to research questions

*RQ1. How much of the recently published scientific literature is freely available according to the data available in Google Scholar, by year of publication, subject categories, and country of affiliation of the authors?*

Google Scholar provided links to freely available versions of documents indexed in the Web of Science and published in 2009 or 2014 in approximately 54.6% of the cases. The percentage is slightly lower for documents published in 2009 (53%) than for documents published in 2014 (55.8%). However, there are important differences at the subject level and at the country level.

Categories related to the Natural and Health sciences achieve the highest percentages of free availability (Basic Life Sciences: 67.5%; Biomedical Sciences: 62.5%). Categories related to the Social Sciences, excepting Psychology (57.8%) and Economics & Business (55.2%) reach lower percentages (Sociology and Anthropology: 40.7%; Social and Behavioral Sciences, Interdisciplinary: 45.4%; Educational Sciences: 40%). Categories in the Arts and Humanities achieve the lowest percentages (Language and Linguistics: 39.4%; Creative Arts, Culture, and Music: 20.9%; Literature: 14.2%).

At the country level the percentages range from approximately 70% overall availability (Brazil, the Netherlands, Switzerland) to approximately 45% (China, Iran, and Russia), if we consider the top 25 countries with a higher output.

These results are remarkably similar to the ones found in other recent large-scale studies that analyse similar datasets but use different mechanisms to find evidence of OA (Piwowar et al., 2018; Science-Metrix Inc., 2018; van Leeuwen et al., 2017).

*RQ2. How much is openly accessible in a sustainable and legal way, and what proportion is freely available but does not meet these criteria?*

We consider that sustainability and legality in OA is important from a policy perspective. For this reason in this study we made a distinction between what we considered reasonably sustainable and legal sources (publishers and repositories), and sources that did not meet these criteria (academic social networks, personal websites, harvesters, and other websites).

Considering the two publication years under study (2009 and 2014), only 33.9% of the documents are openly accessible from sustainable and legal sources. This percentage is formed by the sum of all forms of OA provided by the publisher (Gold, Hybrid, Delayed, and Bronze: 23.1%), and OA provided by repositories that is not also available from the publisher (Green only: 10.8%). Bronze OA is the most common form of OA provided by the publishers. 13.2% of all documents in our sample were available as Bronze OA, while the combination of Gold, Hybrid, and Delayed only made up for 10.1% of the total number of documents. In the Bronze variety of OA, no Open License is available, and publishers usually extend very few rights to the user apart from free access. Therefore, Bronze OA articles cannot be redistributed or reused by anyone without explicit permission from the publisher, thus introducing a legal restriction in the OA consideration of Bronze OA publications.

As for Green OA, 17.6% of the documents in our sample were available from repositories according to Google Scholar.

Using Google Scholar as source of data made it possible to detect that 40.6% of the documents in our sample are freely available from sources that are not considered to meet the criteria of sustainability and legality. This means that more documents are freely available in unsustainable sources and/or in violation of their copyright, than through sustainable and legal ways. In addition to that, 20.7% (of all the documents in our sample) are only freely available from these other sources.

*RQ3. What is the distribution of freely available documents by web domains?*

As other studies had previously hinted (Jamali & Nabavi, 2015; Martín-Martín et al., 2014), the main source of freely available documents according to Google Scholar is, by far, the academic social network ResearchGate, which provided free access to 32.6% of all the documents in our sample (almost the same amount as all publishers and repositories put together). ResearchGate has a strong presence in Google Scholar, demonstrated by the fact that Google Scholar selects the ResearchGate version of an article as the primary version (see Figure 2) in 43.8% of the cases.

After ResearchGate, among the first places of the rank of websites that provided more freely available documents we can find the repositories PubMed Central and arXiv, the academic social network Academia.edu, harvesters like CiteSeerX and Semantic Scholar. After those, we find the largest commercial publishers (Wiley, Elsevier, PloS, Springer-Nature, BioMed Central, Hindawi, MDPI, and Taylor & Francis). In the majority of the cases when there is a freely accessible version of a document from the publisher, Google Scholar selects that version as the primary version.

## 5.2. Final remarks

From the answers to the research questions posed by this study, some general remarks can be drawn about the current status of OA to scientific publications:

The data available in Google Scholar, combined with the data available in other open resources such as CrossRef, DOAJ, OpenDOAR, and ROAR, can provide a faithful representation of OA levels of scientific publications. The results obtained with Google Scholar are similar to other existing approaches of OA identification (e.g. Unpaywall, Science Metrix or van Leeuwen's) thus suggesting some degree of agreement among the different approaches depending on how OA and FA are defined. However, as long as the data available in Google Scholar is not made available to the scientific community, Google Scholar cannot be considered a viable option to analyze OA levels on a regular basis. That said, the fact that Google Scholar, currently the most widely used academic search engine, is able to direct users to freely available versions of documents even when they are not freely accessible from the publisher or from repositories, is something that should not be ignored if one is to truly understand how scientific information is being accessed throughout the world nowadays. Unpaywall can be seen as a strong alternative to find only legal sources of OA, although future research should focus on how the concurrence of several methods could help to depict the most exhaustive landscape of multiple and diverse forms of OA (and FA).

Regarding the prevalence of the different variants of OA, this study confirms that most of the documents that publishers make freely accessible (e.g. Bronze OA) do not specify a clear OA-compatible license. Although this category might contain some masked Gold, Hybrid, or Delayed OA because of practical limitations (see section 4.1 above), it is likely that most of the documents categorised as Bronze OA were intended to be released by the publishers as Gratis Access. If this is the case, it would mean that continued free access over time to a large fraction of documents is entirely dependent on the publishers. This is a precarious situation, because even if publishers' original intention is to maintain Gratis Access status in perpetuity, as sole copyright holders nothing could stop them if they decided to revoke that status in the future. Moreover, in the best-case scenario (where Gratis Access status is maintained over time), the rights extended to users in these cases are very limited (for example, no redistribution and/or limited or no reuse rights), far from what the BOAI initially envisioned. This situation calls for a discussion among all stakeholders regarding the minimum requirements for OA status in scientific articles. But, even if an agreement is not reached, policy makers and funders should still strive to be clear in their OA mandates about the specific accessibility criteria that the outputs of research done with their funds should meet.

As for the OA levels at the country level, this study shows that even though many high-output European countries have OA levels that are above the world average, most of them are still far from complete OA adoption. This means that the goal of reaching 100% OA of scientific publications by 2020 proposed by the European Union in 2016 (Enserink, 2016) is probably unrealistic for most EU countries.

As other studies previously found (Borrego, 2016), the results of this study suggest that even with the current limitations that publishers impose on self-archiving (Gadd & Troll Covey, 2016; Tickell et al., 2017), there is much room for the growth of Green OA, because most publishers do not set limitations for archiving preprints, and some allow the archiving of author's accepted manuscripts at least in some types of websites with no embargo. However, the reality is that many authors still do not do this. What's more, this study confirms that when authors self-archive their documents, they vastly prefer ResearchGate over repositories. ResearchGate has succeeded in convincing researchers from all fields and all over the world to upload massive amounts of documents to its platform, something that institutional repositories have not managed to do. This matches the findings by Borrego (2017) for a sample of Spanish universities. There are several reasons that may have motivated this: the added-value services that ResearchGate provides (e.g. automatically updating the profiles of researchers, the easiness to upload publications, detailed impact and usage indicators that allow the 'quantification of the self' (Hammarfelt, de Rijcke, & Rushforth, 2016; Orduna-Malea, Martín-Martín, & Delgado López-Cózar, 2016), etc.), the prominence with which documents hosted in ResearchGate are displayed in Google Scholar (which might have served as a way to introduce users to the platform), the lack of awareness by researchers of the existence of repositories at their institutions, ignorance on how to use them, usability problems, the increasing barriers to self-archiving imposed by publishers (by which, unlike ResearchGate, repositories usually abide), as well as the lack of academic incentives for scholars to self-archive their work, in opposition to the "immediate feedback and gratification" provided by these academic networks (Hammarfelt et al., 2016). Whatever the reasons, this presents a problem for the advancement of a sustainable and legal system of OA and Open Science in general, because researchers are dedicating their efforts to feeding a proprietary platform that does not make its data available to the scientific community and which may disappear the moment it is not considered profitable.

Lastly, this study confirms that article metadata that contains license information is still not readily available for many articles, making it difficult to categorize the various variants of OA accurately. The appearance of the term "Bronze OA" (Piwowar et al., 2018), which is likely a mix of different variants of publisher-provided OA (Gratis, Gold, Hybrid, Delayed) that cannot be correctly identified because of the lack of license metadata, is a testament of this. CrossRef, currently the largest open source of license information at the article level, strongly recommends publishers to deposit license information, but they are not required to fill this field when they deposit metadata about an article. The system would benefit from the implementation of a standard metadata scheme that defines the specific rights that the license of an article extends to users. This should include the cases of Gratis Access provided by the publishers. This would be a way for publishers to declare their commitment to provide sustainable free access to these articles. In the cases of non-OA documents, self-archiving policies should also be recorded at the article level in machine-readable form, specifying how (under which license), when (specific date for the end of embargo period), where (in what kind of websites), and in what form (preprint, author's accepted manuscript, or version of record) an article can be self-archived. Among other things, this would allow funders and policy makers to check whether a published article meets the terms of a specific OA mandate, and it would allow institutional repositories to monitor the status of the documents published by its researchers more efficiently, and to automate the public release of these documents from the repository under the conditions specified by the license of each article.

In fact, the system suggested above would provide the same functionality as the automated system that the International Association of Scientific, Technical and Medical Publishers (STM) offered to implement in a letter they sent to ResearchGate (STM, 2017). This letter was an attempt to make the social network agree to check for copyright compliance when users upload documents to the platform. However, according to an undated STM announcement, ResearchGate rejected this offer (STM, n.d.), forcing publishers to continue issuing takedown notices when they detect that documents are made freely available in violation of their copyright (Coalition for Responsible Sharing, 2017b). As far as we know, the system has not been mentioned in public again after this exchange, despite its potential usefulness for repository managers all over the world, who, unlike ResearchGate, are usually willing to comply with copyright during the process of deposit.

# Acknowledgements

Alberto Martín-Martín enjoys a four-year doctoral fellowship (FPU2013/05863) granted by the Ministerio de Educación, Cultura, y Deportes (Spain). Funding from the South African DST-NRF Centre of Excellence in Scientometrics and Science, Technology and Innovation Policy (SciSTIP) is also acknowledged. Two anonymous reviewers are also acknowledged for their helpful comments.33